\documentclass[10pt,conference]{IEEEtran}

\usepackage{booktabs} 
\usepackage[flushleft]{threeparttable}
\usepackage{xcolor}
\usepackage[utf8]{inputenc}
\usepackage{float}
\usepackage{amssymb}
\usepackage{amsmath}
\usepackage{ifthen}
\usepackage{caption}
\usepackage[bookmarks=false]{hyperref}
\usepackage{url,moreverb,xspace}
\usepackage[most]{tcolorbox}
\usepackage[dvipsnames,table]{xcolor}
\usepackage{enumitem}
\usepackage{array,graphicx}
\usepackage{soul}
\usepackage{balance}
\usepackage{pifont}
\usepackage{nicefrac}
\usepackage{mathtools}
\usepackage{tabularx}
\usepackage{xcolor,pifont}
\usepackage{tikz}
\usepackage{svg}
\usepackage{pdfpages}
\usepackage{array}
\usepackage{subcaption}
\usepackage{microtype}
\usepackage{todonotes}
\newcommand*\colourcheck[1]{%
  \expandafter\newcommand\csname #1check\endcsname{\textcolor{#1}{\ding{52}}}%
}
\newcommand*\colourcross[1]{%
  \expandafter\newcommand\csname #1cross\endcsname{\textcolor{#1}{\ding{56}}}%
}
\usepackage[vlined, boxruled, linesnumbered] {algorithm2e}
\SetKw{KwBy}{by}
\SetKw{KwBreak}{break}
\SetKw{KwReturn}{return}
\def\HiLi{\leavevmode\rlap{\hbox to \hsize{\color{gray!35}\leaders\hrule height .8\baselineskip depth .5ex\hfill}}}

\newcommand{\davetwo}{\mbox{DAVE-2}\xspace} %

\SetKwComment{Comment}{$\triangleright$\ }{}
\SetCommentSty{itshape}
\newboolean{showcomments}
\setboolean{showcomments}{true}
\ifthenelse{\boolean{showcomments}}
{\newcommand{\nb}[2] {
  \fcolorbox{black}{gray!20}{\bfseries\sffamily\scriptsize#1:}
  {\sf\small$\blacktriangleright$\textit{#2}$\blacktriangleleft$}
}
}
{\newcommand{\nb}[2]{}
}

\newcommand{\head}[1]{\noindent\textbf{#1.}}

\newcounter{fcounter}
\newcommand{\curl}[1]{\footnote{\url{#1}}}

\makeatletter
\newcommand{\linebreakand}{%
  \end{@IEEEauthorhalign}
  \hfill\mbox{}\par
  \mbox{}\hfill\begin{@IEEEauthorhalign}
}
\makeatother

\begin{document}

\pagenumbering{arabic} 
\pagestyle{plain}

\title{Coverage-Guided Road Selection and Prioritization for Efficient Testing in Autonomous Driving Systems}

\author{\IEEEauthorblockN{Qurban Ali}
\IEEEauthorblockA{
\textit{University of Milano-Bicocca}\\
Milan, Italy \\
q.ali@campus.unimib.it}
\and
\IEEEauthorblockN{Andrea Stocco}
\IEEEauthorblockA{
\textit{Technical University of Munich, fortiss}\\
Munich, Germany \\
andrea.stocco@tum.de, stocco@fortiss.org}

\linebreakand
\hspace{-1.0cm} 
\IEEEauthorblockN{Leonardo Mariani}
\IEEEauthorblockA{
\hspace{-1.4cm}\textit{University of Milano-Bicocca}\\
\hspace{-1.4cm}Milan, Italy \\
\hspace{-1.0cm}leonardo.mariani@unimib.it}
\and
\IEEEauthorblockN{Oliviero Riganelli}
\IEEEauthorblockA{
\textit{University of Milano-Bicocca}\\
Milan, Italy \\
oliviero.riganelli@unimib.it}
}

\maketitle

\begin{abstract}
Autonomous Driving Assistance Systems (ADAS) rely on extensive testing to ensure safety and reliability, yet road scenario datasets often contain redundant cases that slow down the testing process without improving fault detection. To address this issue, we present a novel test prioritization framework that reduces redundancy while preserving geometric and behavioral diversity. Road scenarios are clustered based on geometric and dynamic features of the ADAS driving behavior, from which representative cases are selected to guarantee coverage. Roads are finally prioritized based on geometric complexity, driving difficulty, and historical failures, ensuring that the most critical and challenging tests are executed first. We evaluate our framework on the OPENCAT dataset and the Udacity self-driving car simulator using two ADAS models. On average, our approach achieves an 89\% reduction in test suite size while retaining an average of 79\% of failed road scenarios. The prioritization strategy improves early failure detection by up to 95$\times$ compared to random baselines. 
\end{abstract}

\begin{IEEEkeywords}
\itshape Autonomous Driving, Test Prioritization, Deep Learning Testing.
\end{IEEEkeywords}

\section{Introduction}\label{sec:introduction}

The validation of ADAS is premised on the ability to expose failure-inducing behaviors before deployment on public roads~\cite{Cerf:2018:CSC:3181977.3177753}.
Current ADAS pipelines rely heavily on large-scale simulations or log replays, executing thousands of kilometers of driving data to assess driving policy robustness across diverse road conditions.~\cite{Cerf:2018:CSC:3181977.3177753,survey-automotive-outsourcing-analysis,yurtsever2020survey,grigorescu2020survey}. These datasets encompass a wide variety of road topologies, each carrying the potential to trigger critical misbehavior. However, the assumption that more data inherently leads to better validation has created a serious bottleneck: exhaustive replay is computationally expensive, frequently redundant, and offers no guarantee that high-risk cases will be encountered early. As ADAS models scale and regulations demand stronger safety evidence~\cite{1400971}, indiscriminate road execution becomes both impractical and ineffective.

This paper focuses on regression testing of ADAS software, with a particular emphasis on test prioritization. Our work builds on the key observation that, in road-based testing, not all test cases carry the same value. Roads that appear visually distinct may still include segments that induce nearly identical driving behavior, offering no new insight into ADAS performance. Existing regression practices~\cite{birchler-sdc-prioritizer,sensodat} overlook these fine-grained redundancies, as they evaluate entire roads as whole test cases rather than analyzing behavioral variation within them. Furthermore, road execution is often performed in an arbitrary order or guided only by static topology similarity, without considering ADAS behavioral factors. As a result, critical failures are frequently discovered late, while early testing time is spent replaying roads containing benign or redundant segments.

\begin{figure*}[ht]
\centering
\includegraphics[width=0.7\textwidth]{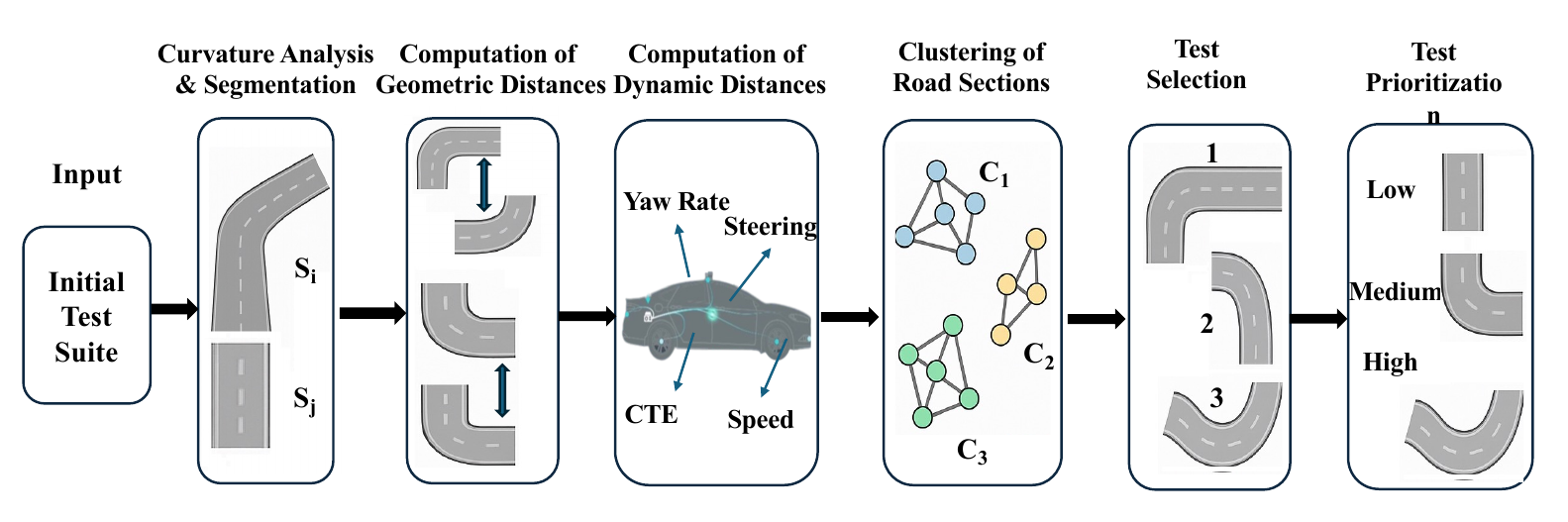}
\caption{Coverage-based selection and prioritization framework.}
\label{fig:workflow}
\end{figure*}

In this work, we introduce a behavior-aware road prioritization framework for ADAS testing, which selects and reorders road execution based on how challenging each segment is for the driving model. In a nutshell, instead of treating all roads equally, our framework directs testing toward those segments most likely to expose failures.
Our framework constructs a behavioral profile for each road derived from trajectory-level signals such as steering variability, oscillatory corrections, cross-track error, and historical instability. Based on these behavioral profiles, we cluster redundant roads, eliminate those offering no novel insight, and rank the remaining segments according to their likelihood of revealing failures. A test case is considered a failure if the car goes Out Of Bounds (OOB), i.e., off the road~\cite{gambi2019automatically}.
We validate our framework using state-of-the-art imitation learning ADAS, including \davetwo and Chauffeur across multiple driving environments, capturing diverse road geometries and driving dynamics across 36 campaigns from the \textsc{OpenCat} dataset~\cite{2025-Ali-ICSEW}. 
The evaluation shows that our prioritization framework consistently identifies the first failure significantly earlier than baseline strategies, including random replay (up to 95$\times$ improvement in the best case, median 78× across campaigns) and geometry-based ordering, and overall execution effort was reduced by more than half (achieving APFD of 0.9 vs 0.2 for random ordering), without affecting failure exposure.

Our study highlights a broader implication for ADAS regression testing: evaluation should be driven by behavioral value rather than only dataset completeness. By focusing on behavioral difficulty, developers can accelerate risk discovery, reduce computational effort, and produce stronger evidence of system safety.

The contributions of this paper are as follows: 

\head{Testing Framework} We introduce a novel behavior-aware framework for prioritizing road-based tests in autonomous driving, moving beyond structural diversity to behavioral challenge. We propose a redundancy filtering method to identify behaviorally equivalent roads and eliminate low-value replay without sacrificing failure exposure.

\head{Evaluation} We conducted a comprehensive empirical evaluation showing substantial improvements in time-to-failure and testing efficiency over conventional replay strategies.

\head{Replication} We provide a replication package containing our implementation and experiment scripts for both case studies to support reproducibility~\cite{replication}.

\section{Background}\label{sec:background}

\emph{ADAS Testing.}
ADASs depend on extensive testing to ensure safe and reliable behavior across diverse real-world conditions~\cite{10-million-miles,Cerf:2018:CSC:3181977.3177753}. Unlike conventional software, where correctness is verified through deterministic inputs and outputs, ADAS behavior emerges from continuous interaction with road geometry, vehicle dynamics, and control policies~\cite{2020-Riccio-EMSE}. Testing must therefore account for dynamic responses such as steering corrections, lane keeping, braking patterns, and the vehicle's ability to remain stable under different conditions~\cite{2020-Humbatova-ICSE,2020-Stocco-ICSE}. 

In current development pipelines, ADAS testing relies heavily on simulation and large-scale road log replay~\cite{2020-Haq-ICST,DBLP:conf/icse/Haq0B22,2023-Stocco-EMSE}. These road datasets are designed to cover a wide spectrum of driving environments. However, exhaustive replay of all available roads quickly becomes infeasible as datasets grow in size. Many roads exhibit redundant behavioral responses, offering minimal new insight into ADAS capabilities. To this aim, test prioritization is essential, as it enables the early execution of tests that are most likely to expose failures~\cite{khatibsyarbini2018test, ali2024testing}.

\emph{Test Prioritization.} Test prioritization is traditionally concerned with ordering test cases so that those most likely to reveal faults are executed earlier~\cite{10.1002/stv.430}. In classical software engineering, prioritization strategies often rely on coverage metrics, change history, or past failures to rank tests. The goal is not to eliminate tests, but to improve the efficiency of fault discovery under resource constraints. Early execution of high-value tests reduces time-to-detection and accelerates developer feedback, which is critical in iterative deployment cycles. 

Transferring this principle to ADAS introduces unique challenges~\cite{birchler-sdc-prioritizer}. Road tests are not isolated function calls; they represent continuous trajectories over time, influenced by control dynamics rather than discrete logic. As a result, conventional source-based prioritization techniques, such as statement coverage or mutation impact, are not applicable as they fail to capture what makes a road test valuable~\cite{2020-Riccio-EMSE}. Instead, value must be derived from behavioral evidence: a road segment that induces high steering oscillation or lateral instability carries more fault potential than a visually complex but behaviorally trivial one~\cite{2018-Arash-ICSE}.

We conjecture that effective prioritization for ADAS requires a shift from visual or input-based diversity toward behavior-based selection. Roads should not be distinguished by how different they appear as a whole, but by how different segments influence the driving model. This reframes prioritization around driving challenge, using fine-grained metrics~\cite{2018-Arash-ICSE} that capture difficulty, instability, or the potential for misbehavior, more effectively than uniformly replaying the entire dataset.
\section{framework} \label{sec:framework}

Our framework includes a pipeline for extracting, comparing, clustering, and prioritizing road sections (called \textit{segments}) from existing whole tests, with the aim of reducing fine-grained redundant test scenarios while preserving the geometric and behavioral diversity of road networks. 

\begin{table*}[!t]
\scriptsize
\centering
\caption{Framework configuration parameters.}
\label{tab:framework_params}
\resizebox{\textwidth}{!}{%
\begin{tabular}{*{5}{l}}
\toprule
\textbf{Parameter} & \textbf{Symbol} & \textbf{Default} & \textbf{Rationale} & \textbf{Sensitivity Range} \\ \midrule
Curvature threshold & $\tau_c$ & 0.015 m$^{-1}$ & Highway design standard for significant curves~\cite{aashto2018policy}  & 0.010–0.025 m$^{-1}$ \\ \addlinespace
Hysteresis window & $w$ & 3 points & Balances noise reduction vs. responsiveness to geometric transitions~\cite{gielis2003generic} & 2–5 points \\ \addlinespace
Min. segment length & $L_{min}$ & 10 meters & Minimum meaningful control sequence for ADAS evaluation~\cite{zhao2019examining} & 5–20 meters \\ \addlinespace
Length ratio threshold & $\tau_{len}$ & 0.8 & Discriminates between direct comparison vs. inclusion matching~\cite{cafiso2012definition} & 0.7–0.9 \\ \addlinespace
Dynamic weight & $w_{dyn}$ & 0.5 & Equal contribution from geometric and dynamic features & 0.0–1.0 \\ \addlinespace
Prioritization weights & $\alpha, \beta$ & 0.5, 0.5 & Equal emphasis on geometric complexity and dynamic difficulty & $\alpha + \beta = 1$ \\ \addlinespace
Historical failure bonus & $H(T)$ & 0.25 & Fixed bonus for previously failed roads & 0.1–0.5 \\ \bottomrule
\end{tabular}
}
\end{table*}

As illustrated in \autoref{fig:workflow}, it starts with curvature-based road segmentation, decomposing roads into different section types (e.g., straight and curved) based on their geometric features. After that, it performs geometric matching using Dynamic Time Warping (DTW)~\cite{Müller2007} to identify repeated patterns across roads while enforcing curvature consistency (left vs right). In addition to geometric features, our framework leverages dynamic driving data (e.g., speed, steering, yaw rate, cross-track error) from simulation logs that supplement the geometric data, enabling more informative and realistic clustering of road sections. 

These combined features jointly drive agglomerative clustering, which groups similar curves with similar ADAS behavior into compact clusters while preserving unique geometries. Once the clusters have been obtained, the next step involves coverage-based road selection, which identifies a subset of roads that collectively cover all clusters. This ensures rare patterns are included while minimizing fine-grained redundancies. 

Finally, the selected roads are ranked higher than the unselected ones, and the roads within each group are ranked based on geometric diversity, dynamic difficulty, and historical failure data, ensuring that the most valuable and challenging scenarios are exercised first. 
Our framework relies on several configurable parameters that influence segmentation granularity, clustering behavior, and prioritization scoring. While we provide recommended default values based on established literature and empirical validation, practitioners may need to adjust these parameters based on their specific testing objectives, available computational resources, and risk tolerance. \autoref{tab:framework_params} presents the key configuration parameters of our framework. 

\subsection{Curvature Analysis and Segmentation}\label{sec:curv_analysis}

\subsubsection*{Initial Test Suite}

The initial test suite $T=\{T_i\}$ consists of tests composed of two main parts $T_i= (R_i, C_i)$, where $R_i$ is the road and $C_i$ is the configuration of the scenario that has to be simulated on the road $R_i$. A \textit{road} is represented as an ordered sequence of Catmull-Rom spline points~\cite{2025-Ali-ICSEW}:
\begin{equation}
\label{eq_road}
R = \{p_1, p_2, \dots, p_n\}, \quad p_i = (x_i, y_i)
\end{equation}
where $(x_i, y_i)$ are the coordinates along the road's centerline. 
The configuration $C_i$ includes the information necessary to run the simulation, such as the initial position of the vehicle and its speed. 

\subsubsection*{Curvature Calculation}\label{sec:curv}

For each road $R$, we calculate the curvature value at each point $(x_i, y_i)$ along the centerline, generating a continuous sequence of curvature values $\kappa_i$ that characterizes the road's geometry. 
Curvature plays a central role in our study, as it captures the essential shape of the road, indicating where it is straight (near-zero curvature values), where it bends (depending on the sign of the curvature value), and how sharply it turns (depending on the absolute curvature value). The curvature calculation employs a \textit{three-point discrete approximation} based on the circumcircle radius of consecutive trajectory points. That is, for three consecutive trajectory points $p_{i-1}=(x_{i-1}, y_{i-1})$, $p_i=(x_i, y_i)$, $p_{i+1}=(x_{i+1}, y_{i+1})$, we first compute the determinant of the consecutive segment vectors to determine the turning direction:
\begin{equation}
\label{eq1}
\det = (x_i - x_{i-1})(y_{i+1} - y_i) - (y_i - y_{i-1})(x_{i+1} - x_i)
\end{equation}
where $\det>0$ corresponds to a counterclockwise (left) turn,  while $\det<0$ corresponds to a clockwise (right) turn, and $\det=0$ indicates collinear points (straight line).
Then the curvature magnitude $\textit{Rad}$ is obtained from the circumcircle radius of the three points~\cite{malgouyres}:
\begin{equation}
\label{eq2}
\textit{Rad} = \frac {|p_i p_{i+1}| \cdot |p_{i-1}p_{i+1}| \cdot |p_{i-1}p_i|} {2|det|}
\end{equation}
The actual curvature is $\kappa = 1/\textit{Rad}$, 
where $\kappa$ = 0 represents a straight line (when $\det = 0$, resulting in $\textit{Rad} \to \infty$).
This method captures both the sharpness of the curve (through curvature magnitude $\kappa$) and direction (through the sign) of road curves. 

To divide roads into smaller, geometrically homogeneous sections for fine-grained focused testing (straight, left curve, right curve), we use segmentation. Segmentation enables fair comparison across roads; without it, comparing two entire roads directly would be misleading, since roads can vary greatly in length and overall shape. Moreover, segmentation highlights distinctive road features and creates a meaningful unit (road section, also called segment) for similarity matching, clustering, and prioritization.

\subsubsection*{Segmentation}\label{sec:segmentation}

The curvature-based segmentation process first employs a \emph{hysteresis-based thresholding} framework~\cite{gielis2003generic} to categorize the shape of the road at each point by also considering the follow-up points, and then creates segments based on the computed information. In particular, given a sequence of $w$ curvature values $\{\kappa_{i}, \kappa_{i+1}, \dots, \kappa_{i+w-1}\}$ starting at point $p_i$ (i.e., $\kappa_{i}$ is the curvature value of point $p_i$), the shape $s_i=\textit{shape}(p_i)$ of the road at position $p_i$ is determined as follows.

\begin{itemize}
    \item \textit{straight}, if $\forall j=1\ldots i+w-1, |\kappa_j| <\tau_c$, 
    \item \textit{left-curve}, if $\forall j=1\ldots i+w-1, \kappa_j >\tau_c$, 
    \item \textit{right-curve}, if $\forall j=1\ldots i+w-1, \kappa_j <-\tau_c$, 
    \item all other cases, retain the classification of the previous point.
\end{itemize}

where $w$ is the length of the window, and $\tau_c$ is the threshold value for straight sections. We use a curvature threshold of $\tau_c = 0.015 m^{-1}$ (corresponding to $R = 66.67 m$) since it has been reported to effectively distinguish between geometrically significant curves that require substantial steering input and nearly-straight sections that can be traversed with minimal control adjustments~\cite{aashto2018policy}. We use a hysteresis window $w=3$ to reduce noise while maintaining responsiveness to genuine geometric transitions, following established principles for discrete curve analysis~\cite{gielis2003generic}.

From a sequence of shape values $s_i$, with $s_i \in \{\textit{straight}, \textit{left-curve}, \textit{right-curve}\}$, segmentation has to establish the boundaries of each section. This is done by extracting subsequences of maximal length $\{s_j, \ldots s_{j+k}\}$ with homogeneous shape, i.e., $s_j=s_{j+i} \forall i=1 \ldots k$ (e.g., a left-curve section where all its points are classified as left-curve). 
A minimum section length constraint of 10 meters is applied to avoid the generation of trivial sections. If a segment below the threshold is extracted, the segment is treated as a noisy section and merged into the preceding section, ensuring that each final section represents a meaningful geometric unit. 
The final output is a \emph{sequence of non-overlapping sections} $S_i$ whose union corresponds to the full road that has been segmented.

\subsection{Computation of Geometric Distances} \label{sec:geometry}

As a result of segmentation, we obtain many road sections, but not all of them are unique. Some sections may repeat across different roads (e.g., straight sections, identical curves) without contributing to exercising additional behaviors of the ADAS. We perform geometric matching to preserve distinctive geometries by identifying redundant sections and filtering out repeated patterns. In particular, we compare sections pairwise using DTW~\cite{Müller2007} on their curvature. Straight sections are flagged for coverage but excluded from similarity matching since they lack distinctive shape patterns.

DTW is a standard algorithm for measuring similarity between two sequences of potentially different lengths by performing non-linear temporal alignment, producing a distance (the difference between two curvature values) that reflects geometric similarity~\cite{Müller2007}. DTW ranges from 0 (identical sections) to 1 (completely different sections). 

Since we look for redundancies in the set of roads to be exercised, we are not only interested in nearly-identical sections but also in the inclusion between segments. In fact, if a section is included in another one, the shorter section would represent a redundant scenario compared to the longer one. To capture this case, we distinguish how we compute the similarity between sections of similar lengths from the matching for inclusion between sections. 

In particular, given two sections, $P$ and $Q$, their length ratio is defined as $\textit{lr}(P,Q)=\frac{min(|P|, |Q|)}{max(|P|, |Q|)}$. If $\textit{lr}(P,Q) <  \tau_{len}$, $sim(P,Q)=1- DTW(P, Q)$, that is, if the two sections have similar lengths, their similarity can be computed by using DTW directly. We empirically determined $\tau_{len} = 0.8$ as a good threshold to discriminate between sections that can be compared directly and sections that must be compared for inclusion in the clusters. 

Suppose the two sections have different lengths (i.e., $\textit{lr}(P,Q) \geq  \tau_{len}$), the similarity between sections $P$ and $Q$ is computed by checking if the shortest section, namely $P$, is included in the longest one, namely $Q$. Note that if $P$ is included in $Q$, using $Q$ for testing implies having already tested the road geometry in $P$, but the opposite is not true.

The inclusion of $P$ in $Q$ is checked by considering every possible alignment between the two sections. In particular, for each alignment position $k$ between 1 and $|Q|-|P|$, the similarity score of the considered alignment is defined as:
\begin{equation}
\label{eq8}
 \textit{sim}_k(P, Q[k:k+|P|]) = 1 - DTW(P, Q[k:k+|P|])
 \end{equation}
 where: $Q[k:k+|P|]$ denotes the subsequence of $Q$ of length $|P|$ starting at index $k$, and $|P|$ is the number of points in $P$.    

Using a sliding-window framework, the shorter section $P$ is aligned at every possible position $k$ along the longer section $Q$, and DTW is computed for each alignment. The final similarity between P and Q is given by the best similarity value computed at various alignment points. This value represents the degree of inclusion of one road into the other, in terms of its geometry. More formally, this is defined as:
\begin{equation}\label{eqSimInclusion}
\textit{sim}(P,Q) = max_{k=1\ldots |Q|-|P|} \textit{sim}_k(P, Q[k:k+|P|])
\end{equation}
Finally, the distance function can be derived from the similarity as follows: $d_{\textit{geom}} = 1-\textit{sim}(P,Q)$.

\subsection{Computation of Dynamic Distances} \label{sec:dynamic}

To better capture scenario difficulty beyond road geometry, we add information about vehicle dynamics to sections and use this information  clustering, creating groups of sections that reflect similarity in both the geometry and driving behavior. 
We extract and store dynamic data when tests are executed for the first time, so that they can be reused for test selection and prioritization when any model change occurs. In particular, we associate each section $S$ with dynamic data $D(S) = \{\textit{di}^S_i\}$ defined as follows: the speed variability $\textit{di}^S_1=\sigma(\{ v_t \}_{t \in S})$, the steering variability $\textit{di}^S_2=\sigma(\{ \theta_t \}_{t \in S})$, the mean cross-track error $\textit{di}^S_3=\tfrac{1}{|S|} \sum_{t \in S} |cte_t|$, and the yaw rate variability $\textit{di}^S_4=\sigma(\{ \dot{\psi}_t \}_{t \in S})$, where ($v_t$ = speed, $\theta_t$ = steering angle, $cte_t$ = cross-track error, $\dot{\psi}_t$ = yaw rate) at time $t$, and $\sigma(\cdot)$ = standard deviation operator.

As these metrics naturally operate on different scales, each feature is normalized to [0, 1] using min-max scaling to make them comparable. The four indicators capture information about the driving difficulty, the control complexity, the correctness, and the stability of the driving~\cite{Anderson2016RAND,Zhao2016Accelerated}.

We compute the distance between the dynamic indicators of two sections $P$, $Q$ as follows.
\begin{equation}
\label{eq:d_dyn}
d_{\mathrm{dyn}}(P, Q) =
\frac{1}{4} \sum_{i=1}^{4}
\left| \textit{di}^P_i - \textit{di}^Q_i \right|
\end{equation}

This yields a distance value in [0, 1], with 0 indicating identical dynamic behavior and 1 maximal dissimilarity.

\subsection{Clustering of Road Sections} \label{sec:clustering}

After section matching and dynamic data integration, we perform clustering to group recurring road patterns based on both geometric and dynamic behavior similarity. This process identifies representative geometries, distinguishing common patterns from rare ones, and ensures that both typical and challenging driving scenarios are included in the test suite. By selecting a minimal set of roads that covers all clusters, we reduce testing redundant segments while maintaining comprehensive road coverage (as explained in \autoref{sec:road_select}). 

The clustering process uses an agglomerative clustering framework~\cite{ward1963hierarchical} organized by section type. 
Agglomerative clustering provides an optimal solution for road test generation by combining automatic cluster determination, hierarchical relationship modeling, deterministic reproducibility, and computational scalability~\cite{ward1963hierarchical, murtagh2012algorithms}. These advantages make it significantly superior to K-means' arbitrary partitioning and non-deterministic behavior~\cite{arthur2007k, jain2010data}, as well as DBSCAN's inappropriate density assumptions and potential exclusion of safety-critical edge cases~\cite{ester1996density}. 
Sections are compared using a distance function that combines distances computed using geometric and dynamic data:
\begin{equation}
\label{eq:hyb}
d_{hybrid}(S_i,S_j) = (1 - w_{dyn}) \cdot d_{geom}(S_i,S_j) + w_{dyn} \cdot d_{dyn}(S_i,S_j)
\end{equation}
with $w_{dyn} \in [0,1]$ (default $w_{dyn}=0.5$).

To identify the groups of similar road sections, we construct a pairwise distance matrix $D$ using $d_{hybrid}$. We then apply the agglomerative hierarchical clustering algorithm with \textit{complete linkage}~\cite{ward1963hierarchical}. It begins by treating each section as an individual cluster, then it iteratively merges the closest clusters until an optimal result is reached (the algorithm automatically determines when to stop according to the distribution of the pairwise distances between elements). Dynamic profiles can become obsolete after significant model changes, which may reduce clustering accuracy. The framework mitigates this through configurable $w_{dyn}$ (geometry-only mode at $w_{dyn}=0$) and hybrid distance metric \autoref{eq:hyb}, ensuring that geometric similarity maintains clustering even when behavioral data is unreliable. For major architectural changes, we recommend a geometry-dominant configuration (e.g., $\alpha = 0.8$, $\beta = 0.2$, $w_{dyn}=0.2$), which minimizes behavioral dependence while preserving selection based on static road properties.

\subsection{Test Selection} \label{sec:road_select}

Once clusters are defined, we identify the representative sections to be covered in each cluster to minimize redundancy while preserving geometric and dynamic diversity. Cluster representatives are used to define coverage requirements, ensuring both typical and challenging scenarios are included. 

We distinguish two cases: singleton clusters and non-singleton clusters.
For each singleton cluster $C=\{s\}$, we select as representative the only section included in $C$, that is $\textit{Rep}(C)=s$. We refer to the set of all the representatives collected from singleton clusters with $\textit{Rep}_{\textit{singleton}}$.
For each cluster with multiple sections, we use a diversity-driven framework to select up to three representatives per cluster. In particular, if the cluster includes three or fewer sections, we select all the sections. If more than three sections are available in a cluster, we compute the mean curvature of each section, and we select three representatives equally distributed across the spectrum of curvature values, intuitively selecting a section with low, medium, and high curvature. 

We indicate this set of representative sections as $R_{\textit{multi}}$.
The selected test cases aim to first cover the representative sections, that is, $\textit{REP} = R_{\textit{singleton}} \cup R_{\textit{multi}}$. 
Once the representative sections $\textit{REP}$ have been identified, the set of tests $T_{\textit{cov}}$ that include all these sections is selected. That is, $T_{\textit{cov}}$ is initialized with the empty set, and then for each section $S \in R$, the test whose road includes $S$ is added to $T_{\textit{cov}}$. 

The output of test selection thus split the initial test suite $T$ into two test suites $T_{\textit{cov}}$, which includes the high-priority test cases, and $T_{\textit{surplus}} = T \setminus T_{\textit{cov}}$, which includes the remaining low-priority test cases.

\subsection{Tests Prioritization} \label{sec:prioritization}
Once $T_{\textit{cov}}$ and $T_{\textit{surplus}}$ have been defined, we determine the execution order of the tests inside each group using a multi-criteria prioritization framework that combines geometric complexity, dynamic behaviour metrics, and historical performance. That is, the tests that use roads with the most complex shape, produce the most challenging ADAS behaviour, and have already failed in past executions are executed first. To achieve this result, given a test $T=(R,C)$, we introduce a test scoring mechanism based on the combination of three scores. 

\begin{figure*}[t]
    \centering
    \includegraphics[width=0.8\textwidth]{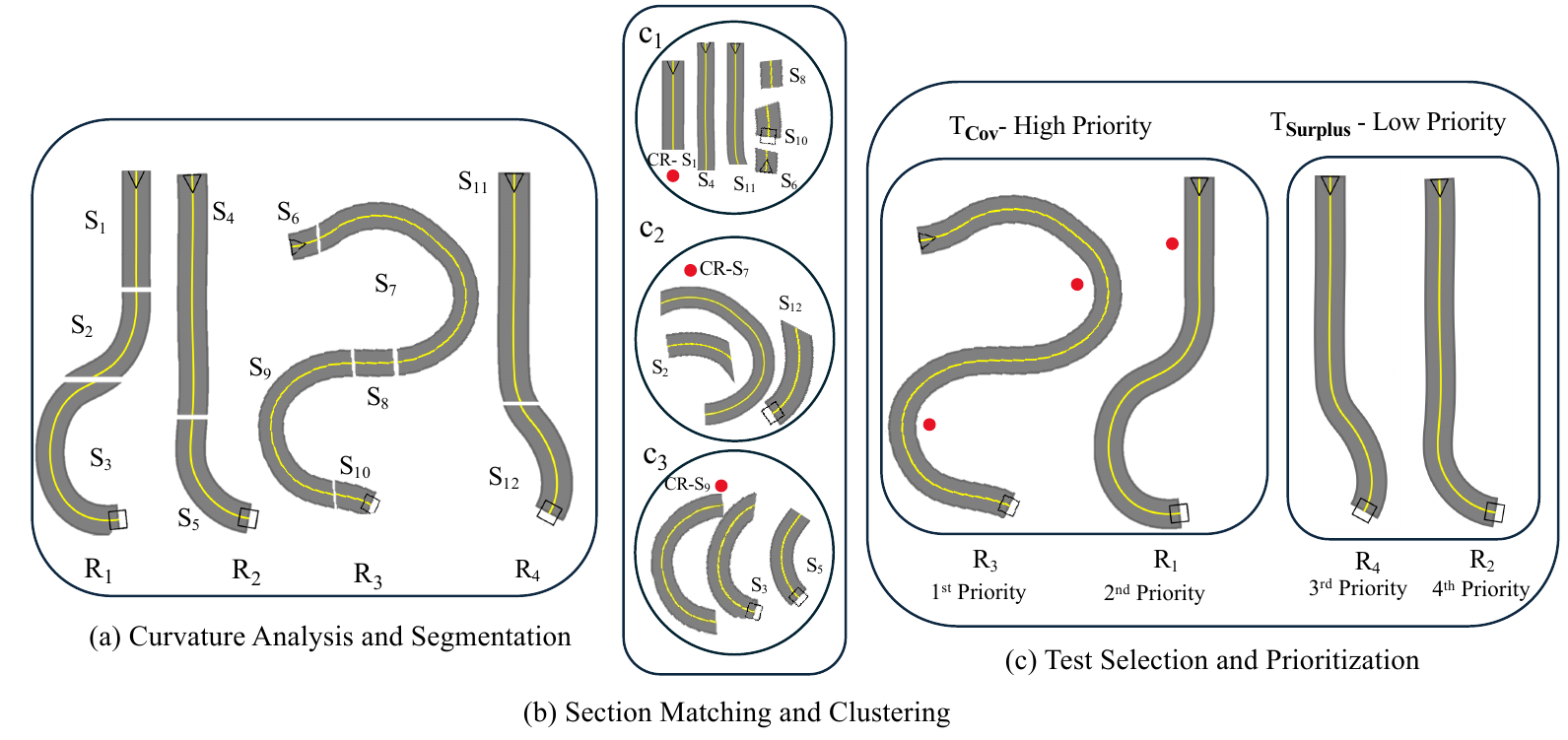}
    \caption{Illustrative example of the proposed framework.}
    \label{fig:example}
\end{figure*}

\head{Geometric score}
Geometric scoring $G(T)$ captures road shape diversity using the metrics curvature variation, number of high-curvature sections, and diversity of section types: 
\begin{equation}
G(T) = w_{cv} \cdot \sigma(\textit{curv}) + w_{hc} \cdot N_{hc} + w_{dt} \cdot D_{types}
\end{equation}

where $\sigma(\textit{curv})$ is the standard deviation of curvature values (curvature variability), $N_{hc}$ is the number of high-curvature sections in $R$ exceeding the threshold $\kappa_{thr}= 0.015$ rad/m, and $D_{types}$ is the number of distinct section types among \{straight, left, right\})($D_{types} \in \{1, 2, 3\}$), with $w_{cv} = w_{hc} = w_{dt} = \frac{1}{3}$ for balanced geometric diversity assessment.

\head{Dynamic score} 
Dynamic scoring $D(T)$ quantifies scenario difficulty as the average of four values: the standard deviation of the speed of the vehicle $\sigma_v$,  the standard deviation of the steering angle $\sigma_\theta$, the mean absolute cross-track error $\text{cte}$, and the standard deviation of the angular velocity $\sigma_{\psi}$.

\head{Historical performance score} 
Historical performance score $H(T)$ contributes by assigning extra points (0.25) to tests that previously caused simulation failures~\cite{rothermel2001prioritizing, zhang2013empirical}.

The final score is a weighted combination:
\begin{equation}
P(r) = \alpha \cdot G(T) + \beta \cdot D(T) + H(T)
\end{equation}
where $\alpha + \beta = 1$, with default $\alpha = 0.5$ (geometric emphasis) and $\beta = 0.5$ (dynamic emphasis).  
All scores are normalized to $[0,1]$ before weighting to ensure comparability. The final execution order inside each group of tests is determined by descending values of $P(r)$.

Our multi-criteria prioritization reflects complementary factors known to influence ADAS failures. Geometric complexity is captured through curvature variation and high-curvature segments, as small-radius and rapidly changing curves are empirically linked to control instability; the curvature threshold $\tau_c = 0.015 m^{-1}$ follows geometric design guidance and prior ADAS studies~\cite{Chang2024evaluation, aashto2018policy}. While geometry captures structural difficulty, dynamic metrics (e.g., steering variability, cross-track error, yaw, and speed variation) capture runtime difficulty beyond geometry and have been shown to be effective for characterizing misbehavior-inducing scenarios~\cite{Zohdinasab2024, Ji2025autonomous}. Finally, historical failure information follows established regression-testing principles, where previously failing or similar tests are more likely to fail again~\cite{rothermel2001prioritizing, Noor2015similarity}. 

The framework assumes that geometric and behavioral features provide complementary signals and that past failures and extreme behaviors are predictive of future risk; however, these signals are often correlated (geometrically complex roads induce challenging behaviors), model and scenario dependent, and unevenly reliable, which can bias prioritization under unseen architectures or environments, highlighting an inherent trade-off between adaptability and generality~\cite{Kim2002history}. Despite these limitations, our empirical evaluation \autoref{sec:empirical-study} shows that the multi-criteria framework consistently outperforms single-criterion baselines (random and geometry-only) across diverse scenarios, indicating that the proposed approach yields more reliable prioritization than relying on any single criterion alone.

\autoref{fig:example} provides a concrete illustration of the proposed framework, demonstrating the sequential operations described in Sections~\ref{sec:curv_analysis}--\ref{sec:prioritization}. 
In Part \textit{(a)}, the input set of roads \( R = \{R_1, R_2, R_3, R_4\} \) is decomposed into multiple sections \( S = \{S_1, S_2, \ldots, S_{12}\} \) through curvature-based segmentation. This process captures local geometric variations such as straight, left-curved, and right-curved sections, enabling fine-grained roads comparison. Although the base taxonomy includes straight, left, and right-curves, higher-order structures (e.g., S-curves, radius transitions) are captured through curvature continuity metrics and segment adjacency, preserving contextual difficulty.

Part \textit{(b)} illustrates the section matching and clustering phase. Each section is compared against others based on its section type (straight vs. straight and so on), leading to the formation of clusters \( C = \{C_1, C_2, C_3\} \). Each cluster group similar sections and is represented by a \textit{cluster representative} (CR), highlighted by a red dot in the figure. The $CR$ serves as the most informative example of that section type (e.g., straight, a sharp curve, or a complex curvature pattern).

Finally, Part \textit{(c)} presents the test selection and prioritization process. Roads containing one or more cluster representatives (e.g., \( R_3 \) and \( R_1 \)) are selected to form the minimal yet diverse test suite, ensuring that the selected set covers the distinct geometric and dynamic behaviors identified across the dataset. These roads are assigned higher priority for testing (e.g., \( R_3 \) first and \( R_1 \) second), while the remaining roads (\( R_4 \) and \( R_2 \)) are retained as lower priority candidates.

 \section{Empirical Study}\label{sec:empirical-study}

\subsection{Research Questions}\label{sec:rqs}

\noindent \textbf{RQ\textsubscript{1} (Selection)}: \textit{How effective is the selected test suite?} 

This RQ investigates the cost-effectiveness of the test suite obtained by selecting the tests that cover the clusters' representatives. A good test suite must discover a good number of issues using a few tests only. 

\noindent \textbf{RQ\textsubscript{2} (Effectiveness of Dynamic Data)}: \textit{To what extent does considering dynamic driving behavior enhance the quality of test selection and the effectiveness of fault detection compared to the geometry-only method?}

This RQ evaluates the individual impact of static data (i.e., road geometry) and dynamic data (i.e., driving behavior) on the overall effectiveness of the test selection and prioritization processes.

\noindent \textbf{RQ\textsubscript{3} (Prioritization)}: \textit{How effective is the proposed prioritization strategy in identifying critical scenarios and detecting failures?}

This RQ investigates whether the proposed prioritization method can schedule test executions in such a way that critical scenarios are likely to be executed early, compared to a random order of the tests.

\noindent \textbf{RQ\textsubscript{4} (Model Transferability)}: \textit{To what extent do test cases selected for one ADAS model reveal failures in architecturally different models?}

This RQ investigates the effectiveness of the prioritization across different model architectures, studying whether the features considered by our approach can transfer across different ADAS models.

\subsection{Objects of Study}

To assess test selection and prioritization, we consider test suites exercising NHTSA~\cite{nhtsa} Level 2 ADAS, which perform vision-based perception tasks using data gathered by the camera sensors of a vehicle. Despite the adoption of Level 2 ADAS in many commercial vehicles, their reliability remains a concern, as evidenced by numerous recent crash reports~\cite{NHTSA-level2-crashes} and real-world validation experiments~\cite{Opletal2025ChinaADAS}. 
Although Levels 3 and 4 ADAS have been proposed~\cite{baiduapolloscapes}, their real-world deployment remains highly constrained. Consequently, addressing the limitations of Level 2 systems is crucial for advancing to higher levels of autonomy. 
We focus on a specific ADAS application, that is, a system for Lane-Keeping Assistance (LKA).

As a model architecture to address the LKA task, we consider \davetwo, which is a convolutional neural network developed for multi-output regression tasks based on imitation learning~\cite{nvidia-dave2}. The model architecture includes three convolutional layers for feature extraction, followed by five fully connected layers. 
\davetwo has been extensively used in a variety of ADAS testing studies~\cite{2023-Stocco-TSE,deeptest,10.1145/3238147.3238187,biagiola2023better,2021-Jahangirova-ICST,biagiola2023boundary}.
The model takes as input an image representing a road scene, and it is trained to predict the vehicle's actuator commands. Our implementation includes a DNN with Lane-Keeping (LK) and Adaptive Cruise Control (ACC) capabilities, as \davetwo is trained to conduct the vehicle on the right lane of the road at the maximum possible speed, by predicting appropriate steering and throttle commands for driving. 

\subsection{Experimental Platforms and Benchmarks}

We conducted experiments on the Udacity~\cite{udacity-simulator} simulation environment since it is open-source and suitable for Level~2 ADAS evaluation. 
Simulation platforms are widely used for testing of ADAS, as researchers have shown that model-level testing is inadequate at exposing system-level failures~\cite{2020-Haq-ICST,briand-offline-emse,2023-Stocco-EMSE}.
Udacity~\cite{udacity-simulator} is developed with Unity 3D~\cite{unity}, a popular cross-platform game engine, based on the Nvidia PhysX engine~\cite{PhysX}, featuring discrete and continuous collision detection, ray-casting, and rigid-body dynamics simulation. 

As a benchmark, we used the test scenarios available in the \textsc{OpenCat} dataset~\cite{2025-Ali-ICSEW}, which provides 32,580 different road scenarios converted from the \textsc{SensoDat} benchmark~\cite{sensodat} across three campaigns: \textit{Ambiegen, Frenetic, and Frenetic\_v}, each generated by a different BeamNG.tech test generator and reflecting distinct testing philosophies. \textbf{AmbieGen}~\cite{Humeniuk2022AmbieGen} applies a multi-objective NSGA-II approach~\cite{Deb2000NSGAII}, optimizing diversity and fault-revealing capability to generate OOB-focused test cases, and \textbf{Frenetic}~\cite{sbst2021} uses a genetic algorithm to minimize the distance between the SDC and the road edge, while \textbf{Frenetic\_v}~\cite{sbst2022} extends Frenetic by reducing invalid roads (e.g., sharp turns or self-intersections), following the SBST Tool Competition road validity definition~\cite{Gambi2022SBST}. 

\subsection{Procedure and Metrics}\label{sec:approach}

\subsubsection{RQ\textsubscript{1} (Selection)}

We evaluate the effectiveness of coverage-based test selection in comparison to random test ordering using the following metrics:

\noindent \head{Reduction Ratio}
The reduction ratio quantifies the efficiency of the test selection process by measuring the percentage of tests eliminated from the original test suite. A higher ratio indicates more efficient reduction. 

\noindent \head{Failed Roads Retention}
This metric measures the effectiveness of the reduction by considering the percentage of failed tests in the selection. We consider the average number of failed roads discovered by a random selection of $N$ tests, for multiple values of $N$, as a baseline.

\subsubsection{RQ\textsubscript{2} (Effectiveness of Dynamic Data)}

To investigate this question, we compare two configurations of the selection pipeline: 
\emph{Geometric-only selection (Geo-Only)}, which clusters and selects tests only using information about the geometry of the road, as described in \autoref{sec:geometry}; and \emph{Hybrid (geometric + dynamic)}, which uses both geometric and dynamic features as described in \autoref{sec:dynamic}. 

Both configurations are applied to the same set of road scenarios across the campaigns, and results are compared in terms of their ability to retain historically failed roads within the reduced subset. We use the same metrics used for RQ\textsubscript{1} to evaluate the effectiveness of the test selection and prioritization. In addition, we measure the relative improvement that dynamic data brings to the use of static data only. 

\subsubsection{RQ\textsubscript{3} (Prioritization)}

We use the following metrics to evaluate the performance of the prioritization approach.

\head{Early Fault Detection (EFD)}
This metric measures the percentage of failed tests selected among the first $k$ tests in the prioritized list. We use this metric to evaluate early fault detection capability, quantifying how quickly our approach identifies safety-critical scenarios compared to random ordering. We consider two cases, $k=10$ to assess the capability to observe misbehaviors immediately after the test suite is executed, and $k=|T_{\textit{cov}}|$ (which is the same number of tests selected by our approach), to assess how good the initial part of the test suite is. 
When considering random ordering, we compute the actual average percentage of failed tests that occur in the first $k$ tests.

\head{Average Percentage of Fault Detection (APFD)}
We compute how good a test ordering is using the well-established APFD~\cite{Rothermel1999TestCP} metric: 
\[ APFD = 1 - \frac{\sum_{i=1}^{m} TF_i}{n \times m} + \frac{1}{2n}
    \]
where $TF_i$ represents the position of the first test revealing fault $i$, $n$ is the total number of tests (roads), and $m$ is the total number of failures. 
Higher APFD values indicate better fault detection across the test execution sequence. 

\begin{table}[t]
\scriptsize
\centering
\caption{Coverage-based selection and prioritization.}
\setlength{\tabcolsep}{1pt}
\resizebox{0.5\textwidth}{!}{
\begin{tabular}{*{10}{c}} 
\toprule

\textbf{Campaign} & \textbf{Total No.} & \textbf{No. Failed} & \textbf{Selected} & \textbf{Reduction} & \textbf{FRR} & \textbf{EFD} & \textbf{EFD10} & \textbf{EFD10} & \textbf{APFD} \\
 & \textbf{Tests} & \textbf{Tests} & \textbf{Tests} & \textbf{\%} & \textbf{Selected\%} & \textbf{RnD \%} & \textbf{Tests} & \textbf{Rnd Tests} & \textbf{} \\

\midrule
\multicolumn{10}{c}{\textbf{Ambiegen Campaigns}} \\
\midrule
\rowcolor{blue!15}  2 & 973 & 11 & 147 & 85\% & 45\% & 0.17\% & 45\% & 1.04\% & 0.92 \\
\rowcolor{blue!15}  3 & 964 & 9 & 206 & 79\% & 89\% & 0.20\% & 80\% & 1.04\% & 0.95 \\
\rowcolor{blue!15}  4 & 965 & 5 & 178 & 82\% & 80\% & 0.10\% & 80\% & 1.04\% & 0.93 \\
\rowcolor{blue!15}  5 & 958 & 10 & 167 & 83\% & 80\% & 0.18\% & 70\% & 1.04\% & 0.91 \\
\rowcolor{blue!15}  6 & 959 & 9 & 179 & 81\% & 78\% & 0.18\% & 70\% & 1.04\% & 0.89 \\
\rowcolor{blue!15}  7 & 963 & 10 & 197 & 80\% & 70\% & 0.21\% & 60\% & 1.04\% & 0.96 \\
\rowcolor{blue!15}  8 & 952 & 11 & 176 & 82\% & 91\% & 0.21\% & 91\% & 1.05\% & 0.92 \\
\rowcolor{blue!15}  9 & 953 & 4 & 187 & 80\% & 100\% & 0.08\% & 75\% & 1.05\% & 0.97 \\
\rowcolor{blue!15} 10 & 971 & 18 & 176 & 82\% & 89\% & 0.34\% & 56\% & 1.03\% & 0.85 \\
\rowcolor{blue!15} 11 & 973 & 10 & 190 & 80\% & 80\% & 0.20\% & 80\% & 1.03\% & 0.93 \\
\rowcolor{blue!15} 13 & 954 & 7 & 185 & 81\% & 86\% & 0.14\% & 86\% & 1.05\% & 0.94 \\
\rowcolor{blue!15} 14 & 959 & 8 & 187 & 80\% & 75\% & 0.16\% & 75\% & 1.05\% & 0.82 \\
\rowcolor{blue!15} 15 & 952 & 19 & 206 & 78\% & 63\% & 0.43\% & 53\% & 1.05\% & 0.96 \\

\midrule
\multicolumn{10}{c}{\textbf{Frenetic Campaigns}} \\
\midrule
\rowcolor{green!17}  2 & 928 & 7 & 27 & 97\% & 57\% & 0.02\% & 57\% & 1.08\% & 0.88 \\
\rowcolor{green!17}  3 & 954 & 11 & 41 & 96\% & 73\% & 0.05\% & 73\% & 1.06\% & 0.91 \\
\rowcolor{green!17}  4 & 964 & 12 & 29 & 97\% & 58\% & 0.04\% & 58\% & 1.05\% & 0.92 \\
\rowcolor{green!17}  5 & 945 & 8 & 30 & 97\% & 63\% & 0.03\% & 75\% & 1.06\% & 0.93 \\
\rowcolor{green!17}  6 & 944 & 16 & 33 & 97\% & 44\% & 0.06\% & 44\% & 1.06\% & 0.90 \\
\rowcolor{green!17}  7 & 967 & 14 & 38 & 96\% & 57\% & 0.03\% & 57\% & 1.03\% & 0.97 \\
\rowcolor{green!17}  8 & 952 & 10 & 30 & 97\% & 60\% & 0.03\% & 70\% & 1.05\% & 0.86 \\
\rowcolor{green!17}  9 & 964 & 6 & 37 & 97\% & 67\% & 0.04\% & 67\% & 1.04\% & 0.97 \\
\rowcolor{green!17} 11 & 866 & 11 & 37 & 96\% & 64\% & 0.05\% & 64\% & 1.15\% & 0.94 \\
\rowcolor{green!17} 12 & 956 & 17 & 39 & 96\% & 59\% & 0.07\% & 59\% & 1.05\% & 0.89 \\
\rowcolor{green!17} 13 & 959 & 13 & 34 & 96\% & 54\% & 0.05\% & 54\% & 1.04\% & 0.95 \\
\rowcolor{green!17} 14 & 866 & 11 & 33 & 96\% & 73\% & 0.05\% & 73\% & 1.15\% & 0.97 \\
\rowcolor{green!17} 15 & 870 & 12 & 37 & 96\% & 67\% & 0.06\% & 67\% & 1.15\% & 0.85 \\

\midrule
\multicolumn{10}{c}{\textbf{Frenetic\_v Campaigns}} \\
\midrule
\rowcolor{orange!20}  2 & 944 & 7 & 31 & 97\% & 86\% & 0.02\% & 87\% & 1.06\% & 0.92 \\
\rowcolor{orange!20}  4 & 525 & 3 & 25 & 95\% & 67\% & 0.03\% & 67\% & 1.90\% & 0.87 \\
\rowcolor{orange!20}  5 & 940 & 7 & 21 & 98\% & 100\% & 0.02\% & 100\% & 1.06\% & 0.95 \\
\rowcolor{orange!20}  6 & 764 & 5 & 22 & 97\% & 100\% & 0.02\% & 100\% & 1.31\% & 0.92 \\
\rowcolor{orange!20}  7 & 47 & 0 & 8 & 83\% & - & - & - & - & - \\
\rowcolor{orange!20} 11 & 953 & 8 & 33 & 97\% & 88\% & 0.03\% & 88\% & 1.05\% & 0.90 \\
\rowcolor{orange!20} 12 & 942 & 7 & 27 & 97\% & 71\% & 0.02\% & 71\% & 1.06\% & 0.87 \\
\rowcolor{orange!20} 13 & 951 & 13 & 34 & 96\% & 54\% & 0.03\% & 87\% & 1.05\% & 0.83 \\
\rowcolor{orange!20} 14 & 934 & 9 & 35 & 96\% & 89\% & 0.04\% & 89\% & 1.07\% & 0.90 \\
\rowcolor{orange!20} 15 & 949 & 7 & 33 & 97\% & 86\% & 0.03\% & 86\% & 1.05\% & 0.85 \\

\bottomrule
\end{tabular}
}
\label{tab:coverage_results}
\end{table}

\subsubsection{RQ\textsubscript{4} (Model Transferability)}

To evaluate cross-model transferability, we used the same set of tests obtained from \textit{RQ\textsubscript{1}} but used the different ADAS model called Chauffeur~\cite {chauffeur2016} instead of the \davetwo. The two models differ significantly in their design and represent architecturally distinct approaches to autonomous driving. DAVE-2 is a feedforward network with five convolutional layers and three fully connected layers ~\cite{nvidia-dave2}, while Chauffeur employs a deeper convolutional structure (six layers with dropout and pooling) and critically integrates recurrent LSTM layers to process temporal context. While DAVE-2 is a larger model, Chauffeur's incorporation of memory through its recurrent design provides a strong basis for evaluating the generalization of our road scenarios across fundamentally different model architectures.

We investigate the capability of the selected test suite to reveal failures also for Chauffeur, analyzing the degree of consistency observed for the tests failed by Chauffeur $(F_C)$ and DAVE-2 $(F_D)$. 

\subsection{Results}\label{sec:results}

\subsubsection{RQ\textsubscript{1} (Selection)}

\autoref{tab:coverage_results} summarizes the outcomes of the coverage-based road selection and prioritization experiments conducted on a total of 32,580 test roads across the three \textsc{OpenCat}~\cite{2025-Ali-ICSEW} campaigns: \textit{Ambiegen}, \textit{Frenetic}, and \textit{Frenetic\_v}. Columns \emph{Campaign}, \emph{total No. Tests} and \emph{No. Failed Tests} indicate the id of the campaign, the number of tests available in that campaign, and the total number of tests failed by \davetwo in the campaign, respectively. 
Columns \emph{Selected} and \emph{Reduction} indicate the absolute number and percentage of tests selected by our approach (i.e., the tests in $|T_{\textit{cov}}|$), respectively. Column \emph{FRR Selected \%} reports the percentage of failures that are revealed by the selected test cases. Column \emph{EFD RnD} indicates the average percentage of faults discovered by randomly selecting as many tests as the ones in $|T_{\textit{cov}}|$.  
Columns \emph{EFD10 Tests} and \emph{EFD10 Rnd Tests} show the percentage of failures revealed by the first 10 test cases selected with our approach and randomly, respectively.
Finally, column \emph{APFD} indicates the average percentage of fault detection. 

The results demonstrate that our approach substantially reduces the number of roads while maintaining strong representational coverage of behavioral and geometric diversity. In the \textit{Ambiegen} campaign, the number of selected roads decreased by an average of 81\% compared to the original dataset, while preserving between 70–100\% of the failed roads. Similar trends were observed for the \textit{Frenetic} and \textit{Frenetic\_v} campaigns, achieving average reductions of 96\% and 94\%, respectively. Despite this drastic reduction, the selection maintained coverage of at least 57–100\% of failed roads, indicating that the clustering-based selection retained roads representing critical geometric and dynamic behaviors, which can be used to discover several failures quickly. Such an early discovery of failures can then be backed up by the execution of the remaining prioritized tests to reveal any possible remaining failure, according to the resources available and development strategy (e.g., first fixing the failures revealed by the prioritized tests before running the remaining tests).

\begin{figure}[t]
\centering
\includegraphics[width=0.5\textwidth, height=8cm, keepaspectratio]{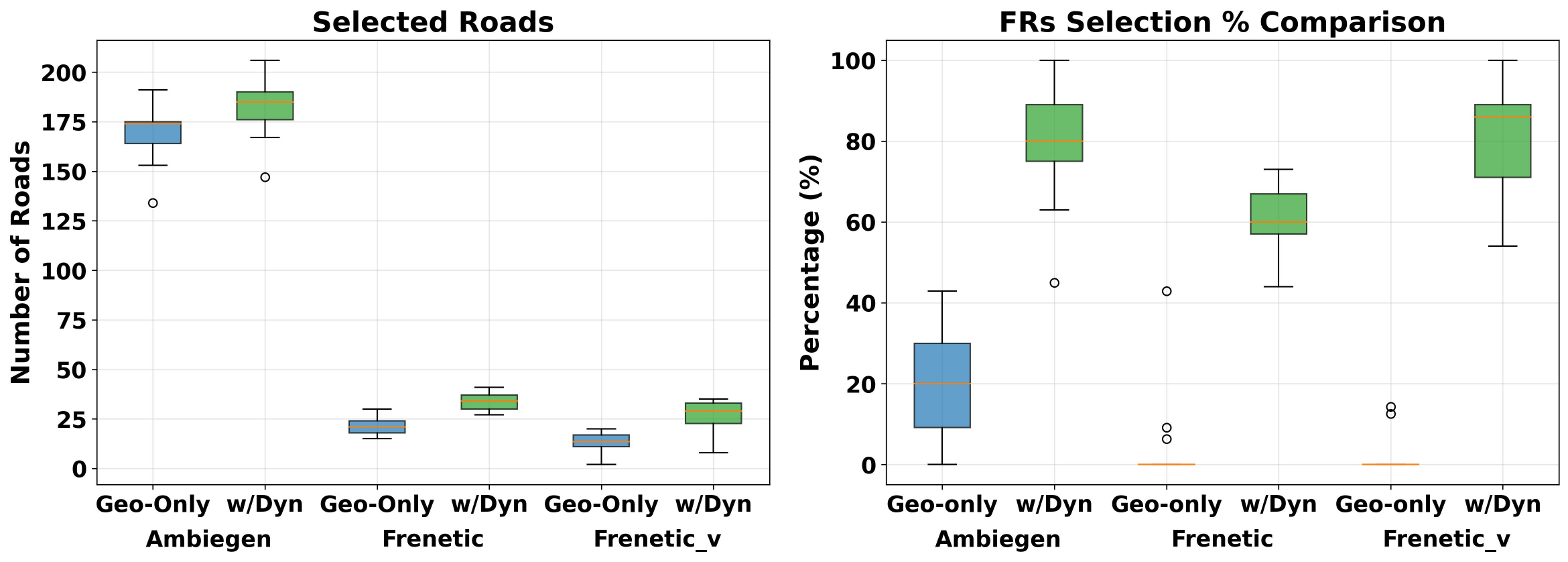}
\caption{\label{fig:geo_vs_dyn} Geometric-only vs Hybrid.} 
\setlength{\belowcaptionskip}{-15pt}
\end{figure}

\subsubsection{RQ\textsubscript{2} (Effectiveness of Dynamic Data)}

\begin{figure*}[t]
    \centering
    \includegraphics[width=0.7\textwidth, height=8cm, keepaspectratio]{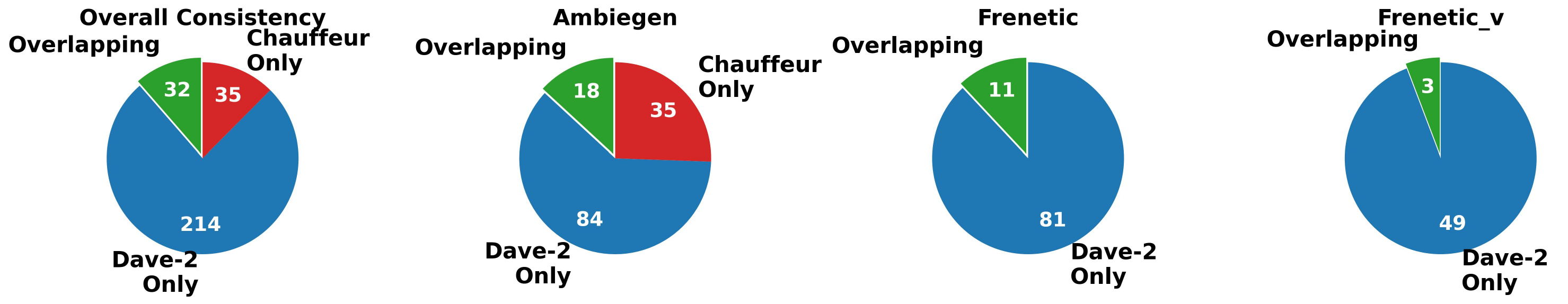}
    \caption{Cross-model test failure consistency analysis for the selected tests.}
    \label{fig:cross_model_con}
\end{figure*}

\autoref{fig:geo_vs_dyn} illustrates the comparative results of test selection (left plot) and fault detection rate (right plot) using road geometry only (\textit{Geo-only}) and including dynamic data (\textit{w/Dyn}). Integrating dynamic behavior data led to substantial improvements in test selection (coverage) and fault detection effectiveness across all campaigns, at the cost of a light increase in the number of selected tests. The slight increase in the selected tests reflects the additional diversity factors introduced by dynamic data. In particular, for the \textit{Ambiegen} campaigns, the average number of selected tests increases from 170 tests (geometric-only) to 185 tests (hybrid) when dynamic attributes were considered, with a corresponding improvement of the fault detection rate from 18\% to 82\%. Similarly,  in the \textit{Frenetic} and \textit{Frenetic\_v} campaigns, the average number of selected roads increases from 20 and 18 to 35 and 32, respectively. Correspondingly, the fault detection rate increases from 5\% and 3\% to 70\% and 65\%, respectively. These results demonstrate that dynamic data is a critical component for test selection.

\subsubsection{RQ\textsubscript{3} (Prioritization)}

\autoref{tab:coverage_results} reports the \textit{EFD} for the selected test suite (column FRR Selected \%) and the same number of randomly selected test cases (EFD Rnd). Moreover, it reports the same metric when only $10$ test cases are selected. This is to study how effective the very top selected test cases are in revealing potential issues in the ADS under test. Specifically, in the \textit{Ambiegen} campaign, the hybrid approach achieved EFD-Top10 rates between 70\% and 86\%, surpassing the baseline ($\approx1.04–1.05\%$). The \textit{Frenetic} and \textit{Frenetic\_v} campaigns exhibited similar trends, where the hybrid method reached up to 87–100\%, while the normalized baseline remained near the statistical expectation ($\approx1.05–1.15\%$). These results demonstrate that the hybrid approach successfully prioritizes the most failure-prone roads early in the testing sequence. Across all campaigns, we achieved approximately 60–90× higher early fault detection efficiency in the hybrid prioritization than the statistical baseline, confirming its effectiveness in ranking safety-critical roads.

In addition to $EFD$, we computed the \textit{APFD} to evaluate the overall prioritization efficiency across the complete test suite. The $APFD$ results range from 0.82 and 0.97 across all campaigns, reflecting the fraction of cumulative fault detection achieved as testing progresses. These values indicate that, on average, 85–97\% of total faults are detected within the first half of the prioritized execution order, demonstrating that faults are concentrated toward the beginning of the sequence rather than uniformly across it. This trend reinforces that the hybrid prioritization approach effectively concentrates high-risk, failure-prone roads early in the execution process, thereby maximizing detection efficiency and resource utilization.

\subsubsection{RQ\textsubscript{4} (Model Transferability)} 

\begin{figure}[!tbp]
    \centering
    \includegraphics[width=0.45\textwidth, height=6cm, keepaspectratio]{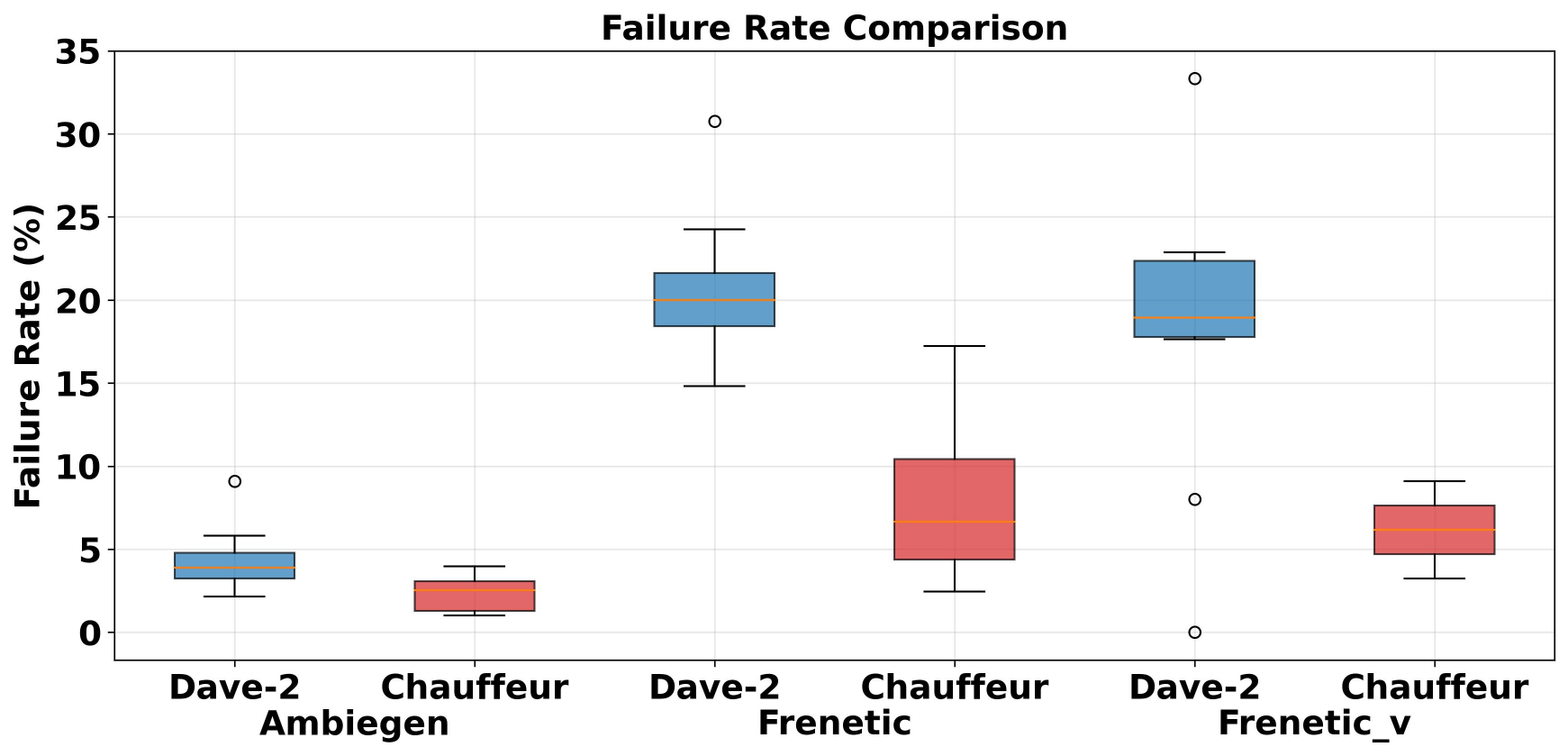}
    \caption{Cross-Model Failure Distribution (Dave-2/Chauffeur).}
    \label{fig:cross_model_dis}
\end{figure}

\autoref{fig:cross_model_con} and \autoref{fig:cross_model_dis} illustrate the results for the cross-model transferability. The cross-model evaluation shows limited transferability of failure patterns between \davetwo and Chauffeur. Using the prioritized test suite 
$T_{\textit{cov}}$ derived from \davetwo behavioral profiles, Chauffeur consistently failed fewer tests than \davetwo across all campaign families, with low to moderate failure overlap (34\% overlap on average). Specifically, in the AmbieGen campaigns, \davetwo and Chauffeur share only 25\% of failures; Frenetic campaigns exhibit higher overlap (52\%), while Frenetic\_v shows minimal overlap (21\%) on average.

These results indicate that failure-inducing characteristics are largely model-specific: test cases stressing \davetwo feedforward architecture do not necessarily challenge Chauffeur LSTM-based controller, and vice versa. Consequently, test suites optimized for one model provide only partial coverage for architecturally different models, with geometric features transferring more reliably than behavioral features. While our framework identifies geometrically complex roads that affect both models, most failures (48-79\%, depending on campaign) remain model-specific. This suggests that, for multi-model testing, practitioners should emphasize geometry-based selection using low dynamic weights ($w_{dyn} =0.0-0.2$), complemented by model-specific behavioral profiling for comprehensive coverage. 

\medskip

Overall, the findings through \textit{RQ\textsubscript{1}-RQ\textsubscript{4}} validate the effectiveness of integrating geometric complexity, dynamic variation, and historical failure information in identifying representative and diverse roads while minimizing redundant tests. The consistent inclusion of most failure-prone roads suggests that the method captures the spectrum of safety-critical geometries and dynamic driving conditions, producing a more discriminative prioritization ranking. Specifically, the findings demonstrate that the proposed selection strategy preserves geometric and behavioral coverage (\textit{RQ\textsubscript{1,2}}), the hybrid prioritization significantly improves early fault detection when a same model is used (\textit{RQ\textsubscript{3}}), and the selected roads have the potential to early reveal failures also in other models (\textit{RQ\textsubscript{4}}). Collectively, these results suggest that the proposed prioritization strategy may generate cost-effective prioritized aware test suites for autonomous driving evaluation.

\subsection{Threats to Validity}

Regarding internal validity, our framework relies on a combination of geometric and dynamic features to guide test selection and prioritization. While the thresholds and weights used were selected based on prior literature and default values, alternative configurations may yield different results. Nevertheless, the consistent performance across multiple campaigns and models suggests that the chosen parameters are robust.  

External validity is limited by the use of the OPENCAT dataset and the Udacity simulator, which offer a rich and diverse set of road scenarios. While these platforms provide a controlled and scalable environment for testing, they may not fully capture the complexity of real-world driving conditions, such as sensor noise, unpredictable traffic, or weather variability. However, the ability to systematically evaluate thousands of scenarios across different campaigns strengthens the generalizability of our approach, and future work will extend validation to physical-world platforms and additional datasets.

Construct validity may be affected by the metrics and features chosen to represent driving behavior and scenario diversity. Although DTW and the selected dynamic features aim to capture meaningful aspects of driving complexity, they may not fully encompass all factors contributing to failure-inducing scenarios, such as rare edge cases or specific driving behaviors. To mitigate this, our study includes multiple metrics and validation across different models and datasets, but further refinement of feature representation and incorporation of real-world noise remains an important direction for future research.


\section{Related work}
\label{sec:related-work} 
ADAS testing faces significant scalability challenges due to the large number of possible driving scenarios and the safety-critical nature of these systems~\cite{lou2022testing, ADS_testing_survey_2022}. As exhaustive testing is computationally infeasible, prior work has focused on intelligent test reduction and prioritization strategies.

Classical test reduction and prioritization techniques aim to maximize fault detection using coverage-based, fault-based, or requirement-based criteria~\cite{yoo2012regression}. However, for autonomous driving systems with vast and highly variable scenario spaces, such techniques often fail to capture behavioral and environmental diversity. Consequently, recent work has explored search-based, scenario-based, and clustering-based approaches that reduce redundancy while preserving diversity and fault detection capability~\cite{deng2022scenario,kerber2020clustering,bernhard2021optimizing,song2023identifying,lu2021search,lu2022learning,birchler-sdc-prioritizer}.

Deng et al.~\cite{deng2022scenario} propose a scenario-based reduction and prioritization approach that segments driving recordings into semantically homogeneous scenes using static and dynamic features, removing redundancy via vector similarity and prioritizing scenarios based on coverage and rarity. While their method operates at the \textit{scene level} using semantic abstractions, our approach focuses on \textit{fine-grained road segments} enriched with geometric and behavioral execution metrics.

Several studies employ clustering to organize large driving datasets. Kerber et al.~\cite{kerber2020clustering} use hierarchical agglomerative clustering based on spatiotemporal features to estimate test coverage in highway scenarios, while Bernhard et al.~\cite{bernhard2021optimizing} apply multi-stage trajectory clustering that combines Gaussian Mixture Models and hierarchical clustering to group vehicle interactions using spatial and directional behavior. Song et al.~\cite{song2023identifying} combine Analytic Hierarchy Process (AHP) with hierarchical clustering to identify critical lane-keeping scenarios, first ranking key failure factors via AHP, then clustering scenarios based on these ranked factors. Unlike these approaches, which focus on \textit{trajectory-level} or interaction-based clustering, our framework performs \textit{road segment-level} clustering that integrates geometric curvature with behavioral dynamics.

Lu et al.~\cite{lu2021search} propose SPECTER, a search-based method that uses multi-objective evolutionary algorithms to balance scenario diversity, criticality, and execution cost. They further introduce DeepCollision~\cite{lu2022learning}, a reinforcement learning–based approach that adapts environmental parameters (e.g., weather, traffic density, obstacle placement) to induce failures during test execution. While these methods excel at \textit{online adversarial} scenario generation during test execution for early failure discovery, our framework targets \textit{offline} selection and prioritization of existing test suites to improve regression testing. Similarly, Birchler et al.~\cite{birchler-sdc-prioritizer} propose SDC-Prioritizer, which leverages static road features to prioritize test cases, demonstrating the effectiveness of geometric features ranking for failure detection.

Overall, existing approaches typically emphasize geometric or behavioral characteristics in isolation, missing opportunities for comprehensive scenario characterization. Our framework addresses this gap by jointly integrating geometric similarity, dynamic behavioral analysis, and historical failure information to enable multi-criteria prioritization of ADAS test suites.
\section{Conclusions}\label{sec:conclusions}

In this paper, we introduced a framework for test selection and prioritization in ADAS, designed to improve testing efficiency while preserving behavioral and geometric diversity. By segmenting road scenarios, incorporating dynamic driving behavior, and clustering to reduce redundancy, the framework selects and orders tests based on their potential to expose failures. 
In our experiments on the OPENCAT dataset with two ADAS models, our approach reduced test suite size by up to 89\% while retaining an average of 73\% of failing cases. The prioritization strategy accelerated fault detection by up to 95$\times$ over random baselines, achieving APFD scores between 0.82 and 0.97, consistently exposing critical failures early. These results demonstrate that the method provides a practical balance between efficiency and robustness, making it suitable for cost-effective ADAS regression testing.

\section{Acknowledgments}
This work has been supported by the Centro Nazionale HPC, Big Data e Quantum Computing (PNRR CN1 spoke 9); the MUR under the grant ``Dipartimenti di Eccellenza 2023-2027'' of the DISCo Department of the University of Milano-Bicocca, and the Bavarian Ministry of Economic Affairs, Regional Development and Energy.


\balance
\bibliographystyle{IEEEtran}
\bibliography{paper}

\end{document}